\documentclass[aps,prb,draft,superscriptaddress,twocolumn,showpacs,intlimits,amsmath,amssymb,floats]{revtex4}
\usepackage{bm}
\usepackage[final]{graphicx}
\usepackage{epsfig}
\usepackage{color}
\definecolor{DarkBlue}{rgb}{0.1,0.1,0.5}
\definecolor{Red}{rgb}{0.9,0.0,0.1}
\definecolor{Green}{rgb}{0.0,0.99,0.0}

\begin{document}

\title{A Fidelity Study of the Superconducting Phase Diagram in the 2D Single-band Hubbard Model}

\author{C.~J.~Jia}
\affiliation{Stanford Institute for Materials and Energy Sciences,
SLAC National Accelerator Laboratory, 2575 Sand Hill Road, Menlo
Park, CA 94025.}
\affiliation{Departments of Applied Physics, Stanford University, CA 94305.}
\author{B.~Moritz}
\affiliation{Stanford Institute for Materials and Energy Sciences,
 SLAC National Accelerator Laboratory, 2575 Sand Hill Road, Menlo
 Park, CA 94025.}
\affiliation{Department of Physics and Astrophysics, University of North Dakota, Grand Forks, ND 58202, USA}
\author{C.-C.~Chen}
\affiliation{Stanford Institute for Materials and Energy Sciences,
SLAC National Accelerator Laboratory, 2575 Sand Hill Road, Menlo
Park, CA 94025.}
\affiliation{Departments of Physics, Stanford University, CA 94305.}
\author{B. Sriram Shastry}
\affiliation{Department of Physics, University of California, Santa
Cruz, CA 95064.}
\author{T.~P.~Devereaux}
\affiliation{Stanford Institute for Materials and Energy Sciences,
SLAC National Accelerator Laboratory, 2575 Sand Hill Road, Menlo
Park, CA 94025.}
\affiliation{Geballe Laboratory for Advanced
Materials, Stanford University, CA 94305.}

\date{\today}

\begin{abstract}
Extensive numerical studies have demonstrated that the two-dimensional single-band Hubbard model contains much of the key physics in cuprate high-temperature superconductors. However, there is no definitive proof that the Hubbard model truly possesses a superconducting ground-state or, if it does, how it depends on model parameters. To answer these longstanding questions, we study an extension of the Hubbard model including an infinite-range $d$-wave pair field term which precipitates a superconducting state in the $d$-wave channel. Using exact diagonalization on 16-site square clusters, we study the evolution of the ground-state as a function of
the strength of the pairing term. This is achieved by monitoring the fidelity metric of the ground-state, as well as determining the ratio between the two largest eigenvalues of the $d$-wave pair/spin/charge density matrices. The calculations show a $d$-wave superconducting ground-state in doped clusters bracketed by a strong antiferromagnetic state at half-filling controlled by the Coulomb repulsion $U$ and a weak short-range checkerboard charge ordered state at larger hole-doping controlled by the next-nearest-neighbor hopping $t^{\prime}$. We also demonstrate that negative $t^\prime$ plays an important role in facilitating $d$-wave superconductivity.
\end{abstract}

 \maketitle




\section{Introduction}

Originally proposed for the study of transition metals, the
Hubbard model\cite{Hubbardinitial} continues to be one of the simplest, and yet complex models, used to describe strongly correlated systems and possibly the cuprate high-temperature superconductors\cite{AndersonZRS,ZhangRice}. However, despite over 40 years of intensive numerical and analytical studies and over 20 years devoted to the connection with superconductivity in cuprates, the repulsive Hubbard model in two dimensions has not been proven to possess a unique superconducting (SC) ground-state in any parameter space apart from asympototically small coupling with a vanishingly small transition temperature \cite{NewMechanicsforSC,Raghu}. Promisingly, recent dynamical cluster approximation calculations suggest the presence of finite temperature $d$-wave superconductivity near half-filling\cite{QuantumClusterTheories,MaierExtra}, yet the true ground-state remains inaccessible due to the fermion sign problem.

Empirically, in terms of the underlying noninteracting band structure, the SC transition temperature is larger in materials possessing bonding Fermi-surfaces far from $(\pi,0)$ with only a small portion nesting the antiferromagnetic reciprocal lattice vector $Q=(\pi,\pi)$\cite{longhopping1,Tanakat',Ref9Aadded,Ref9Badded}. This degree of Fermi surface bending is related to the strength of the hopping parameter along the diagonals of the unit cell - the next-nearest-neighbor hopping $t^\prime(<0)$ - in the single-band model for the electronic dispersion. Pavarini \emph{et al.} give reasons why a large $t^\prime$ (the absolute value) might favor superconductivity
due to the minimization of the hybridization between cooper and apical oxygen orbitals\cite{longhopping1}. However,
 most numerical studies of superconductivity in the single-band
 Hubbard model do not show a strong dependence on $t^\prime$ - in
 fact most studies of the $t$-$J$ and the Hubbard model indicate
 that negative $t^\prime$ hurts
 superconductivity\cite{t'hurts,t'hurtsWhite,ThMaiert',ttJpair,t'hurtsGFMC,t'HubbardHolstein}. Thus, this intriguing clue to the root cause of superconductivity remains unexplained.

  In addition, the ``phase diagram" of materials in the cuprate family has displayed several tendencies to show charge ordering in doped systems, forming either static stripes as revealed in neutron scattering studies of LaBaCuO and related compounds \cite{NeutronScatter,KivelsonRMP}, or local checkerboard-type order on the level of a few lattice spacings as revealed by scanning tunneling spectroscopy on SC Bi$_2$Sr$_2$CaCu$_2$O$_{8+x}$ and Bi$_2$Sr$_2$CuO$_{6+x}$ \cite{Bi2201Hoffman,Bi2201Kapitulnik,Bi2201Pseudo,Bi2201Davis,Bi2201Hudson,STMCheckboard}. Moreover, recent angle-resolved photoemission\cite{Bi2201Makoto,Ref24Added} on nearly optimally doped Bi$_2$Sr$_2$CuO$_{6+x}$ has revealed a density-wave type of order emerging below temperatures $T$* above $T_c$. This propensity for charge ordering in doped systems gives an indication that a density-wave type of instability may also lie in close proximity as a competing phase to the dominant SC order parameter, and it suggests that higher SC transition temperatures might be obtained as a result of controlling the emergence of the density-wave phase. After such a long period of study, it
remains an open question whether this model has enough
physics to capture the essence of the ground-state phases
in the cuprates as a function of doping or other material
parameters.

Our aim is to investigate the conjecture that the Hubbard model could possess a $d$-wave SC ground-state, as well as other postulated phases that either compete or coexist with $d$-wave superconductivity. For this, we employ numerical calculation in small clusters to study an extension of the Hubbard model, including an infinite-range $d$-wave pair field term
of strength $\lambda$, which precipitates superconductivity in the
$d$-wave channel. The method of adding a pair field term to a standard low-energy Hamiltonian was introduced earlier by Rigol, Shastry, and Haas (RSH) for the
study of superconductivity in the $t$-$J$ model\cite{Shastry1,
Shastry2}. Our Hamiltonian guarantees a $d$-wave SC ground-state for sufficiently strong $\lambda$. By gradually decreasing $\lambda$, we expect a quantum phase transition (QPT) or, in a finite-size cluster, a quantum critical crossover (QCC),  provided the Hubbard model does not favor a $d$-wave SC ground-state.
Otherwise, if no QPT or QCC is observed, we infer that the bare Hubbard model possesses a $d$-wave SC ground-state.

We establish a criterion for the existence of the QCC by monitoring the fidelity metric $g$ (Ref.~[\onlinecite{fidelityorigin2,FidelityScaling,Fidelity_Hubbard_analytical,Fidelity_TI_QPT}]) as a function of $\lambda$. We perform exact diagonalization (ED) calculations on the 16B Betts cluster\cite{BettsCluster} (16-site square cluster) for the two-dimensional(2D) single-band Hubbard model near half-filling for different values of the Coulomb repulsion $U$ and the next-nearest-neighbor hopping $t^\prime$. We also look at the strongest competing order that appears in proximity to the SC phase. For this purpose, we study the QCC between $d$-wave SC and other
phases as $\lambda$ gradually decreases by calculating the fidelity metric and the corresponding density matrix eigenvalues and their ratios, as discussed below in Eqs.~(7)-(9).

After we obtain QCC information for the hole-doped region of interest, we plot the phase diagram. As a nontraditional phase diagram, it tracks the critical strength of the $d$-wave SC term, separating the SC dominant phase and other phases, as a function of doping. The phase at $\lambda=0$ is simply the phase of the Hubbard model with no additional pairing term. We emphasize that the $d$-wave SC state in the phase diagram is robust, since this state is favored by the pairing term added to the standard Hubbard model, and the fidelity metric is robust in tracking the phase transition as the pairing term is continuously turned off. We examine the other competing orders in a less robust manner, by calculating the dominant order of each phase via studying the charge and spin density matrices. Although finite-size scaling using larger cluster calculations is currently unfeasible, the small cluster calculations could provide insights into the trends of the behavior of $d$-wave SC and other possible orders approaching the  thermodynamic limit. As one of the most important results in our work, we will present this phase diagram in Sec. IV.

The rest of the paper is organized as follows. In Sec. II we present our calculational methodology and justify the use of the fidelity metric to monitor the change of orders in ground-states. In Sec. III
we present and discuss our results from finite-size cluster ED studies of the single-band Hubbard model for various hole-dopings and other model parameters. Finally, we offer a summary of our results and interpretations of the phase diagram in Sec. IV.

\section{Calculation Methodology}
\subsection{Model Hamiltonian}

The single-band Hubbard model is described by the Hamiltonian
\begin{equation}\label{Hubbard}
\begin{aligned}
\mathit{H_{Hubbard}}=-\sum_{i,j,\sigma} t_{ij}c_{i\sigma}^{\dagger}c_{j\sigma}+\sum_{i} U n_{i\uparrow} n_{i\downarrow},
\end{aligned}
\end{equation}
where $c_{i\sigma}^{\dagger}$ ($c_{i\sigma}$) creates (annihilates) an electron with spin $\sigma$ on lattice site $i$; $t_{ij}$ are overlaps or ``hopping" integrals restricted to the nearest-neighbor and
next-nearest-neighbor, denoted as $t(>0)$ and $t'(<0)$; $U$ is the on-site Coulomb repulsion; and $n_{i\sigma}=c_{i\sigma}^{\dagger} c_{i\sigma}$ is the number operator.

To ensure a $d$-wave SC ground-state, we introduce a
$d$-wave pair field term of the BCS type with a $d$-wave order parameter\cite{BCS}:
\begin{equation}\label{d-wave BCS}
\mathit{H_{d-wave}} = -\frac{\lambda}{N}\sum_{i,j} D_i^{\dagger}D_j,
\end{equation}
where $D_i = \Delta_{i,i+\hat{x}}-\Delta_{i,i+\hat{y}}$, $ \Delta_{ij} =
c_{i\uparrow}c_{j\downarrow}+c_{j\uparrow}c_{i\downarrow}$, $\lambda$ is the strength of this $d$-wave BCS attraction, and $N$ is the total number of lattice sites. We will assume that $\lambda$ changes between $0$ and $O(1)$, in units of $t$.

RSH have shown that for the $t$-$J$ model plus a $d$-wave pair field term with strength $\lambda$, a SC ground-state is present at some sufficiently large $\lambda$ in small clusters\cite{Shastry1,Shastry2}. Similar to this model, our Hamiltonian is assumed to possess a $d$-wave SC ground-state when $\lambda$ is sufficiently large in any finite-size cluster, without loss of generality. Thus, we view $\lambda$ as a ``continuous switch" to study the Hubbard model containing a $d$-wave SC ground-state: when $\lambda$ is large or ``on", the $d$-wave BCS term is dominant and the Hamiltonian exhibits a $d$-wave SC ground-state; as $\lambda$ approaches zero or the ``off" state, the Hamiltonian reduces to that of the bare Hubbard model, thus revealing the properties of its ground-state. This is completely analogous to the study of ground-state magnetism by slowly switching off an applied magnetic field. In summary, we begin with a large $\lambda$, utilize the assumed $d$-wave SC ground-state as a starting point, and gradually decrease $\lambda$. If a QCC does not occur, as indicated by the behavior of the fidelity metric defined in the following section, it illustrates that the Hubbard model ground-state has the same characteristic as the ground-state at large $\lambda$: a $d$-wave SC state; if a QCC occurs, it demonstrates a change in the ground-state phase, indicating that the Hubbard model does not possess a $d$-wave SC ground-state. The nature of the non-SC state is then determined from a study of the density matrix, as elaborated below.

\subsection{Fidelity metric}

The main point of this work, capturing a possible QCC between the $d$-wave SC state and other phases, can be accomplished by calculating a quantity known as fidelity. Fidelity was first used widely in quantum information to provide the criterion for distinguishing between
quantum states\cite{quaninfsci1,quaninfsci2} and has now been utilized increasingly in general physics for investigating classical as well as quantum critical behavior in
various systems\cite{fidelityorigin2,FidelityBoseHubbard,FidelityScaling,Fidelity_on_Hubbard_Model,Fidelity_XY_Model,
Fidelity_TI_QPT,Fidelity_Hubbard_analytical,Fidelity_in_TI_QPM,Fidelity_Susceptibiliy_abnormal}. The work with fidelity on fermionic models and spin models recently has received increased attention related to QPT and is powerful enough for nonstandard QPT like topological phase transitions,\cite{Fidelity_TI_QPT,Fidelity_in_TI_QPM} where no direct examination can be made of an order parameter.

Given a Hamiltonian of the form $H(\lambda)=H_0+\lambda H_1$, the overlap between the ground-states obtained with small changes in the parameter $\lambda$
\begin{equation}
F(\lambda,\delta\lambda) = \langle\Phi_0(\lambda)|\Phi_0(\lambda+\delta\lambda)\rangle
\end{equation}
defines the fidelity. $\lambda$ need not be a small parameter but can be tuned to lie near a critical value such that the ground-state wave functions change abruptly with small changes in $\lambda$ near this critical value.  For small values of $\delta\lambda$, $|\Phi_0(\lambda+\delta\lambda)\rangle$ can be expanded in the normalized eigenstates at $\lambda$:
$|\Phi_{\alpha}(\lambda)\rangle$ ($\alpha = 0, 1, 2...$); to second order in $\delta\lambda$ the term
proportional to the unperturbed ground-state is
\begin{equation}
\langle\Phi_0(\lambda)|\Phi_0(\lambda+\delta\lambda)\rangle = 1-\frac{\delta\lambda^{2}}{2}\sum_{\alpha\neq0}\frac{|\langle\Phi_{\alpha}
(\lambda)|H_1|\Phi_0(\lambda)\rangle|^{2}}{[E_0(\lambda)-E_{\alpha}(\lambda)]^2}
\end{equation}
 The fidelity has been shown to be an extensive quantity that scales with the system size in other Hamiltonians similar to our model\cite{Fidelity_on_Hubbard_Model,FidelityScaling}. So, following RSH, we introduce the fidelity metric\cite{fidelityorigin2,Fidelity_Hubbard_analytical,Fidelity_TI_QPT,FidelityScaling}
\begin{equation}
g(\lambda,\delta\lambda)=(2/N)(1-F(\lambda,\delta\lambda))/\delta\lambda^2 \end{equation}
(equivalent to the fidelity susceptibility defined in other contexts\cite{Fidelity_on_Hubbard_Model,Fidelity_Susceptibiliy_abnormal}).
, whose limit as $\delta\lambda \to 0$ is well defined away from the critical points when perturbation theory applies
\begin{equation}
\lim_{\delta\lambda \to
0}g(\lambda,\delta\lambda)=\frac{1}{N}\sum_{\alpha\neq 0}
\frac{|\langle\Phi_{\alpha}(\lambda)|H_1|\Phi_0(\lambda)\rangle|^2}
{[E_0(\lambda)-E_{\alpha}(\lambda)]^2}.
\end{equation}
From Eqs.~(5) and (6), the fidelity metric is in principle an intensive quantity that does not scale with the number of sites and measures how significantly the ground-state wave function changes as the parameter $\lambda$ changes.
At the critical point where perturbation theory fails, the normalized ground-state wave functions differ significantly with only a small change in the parameter $\lambda$, which results in a significant decrease of the overlap of these two wave functions or the fidelity at that point. Thus, we expect the fidelity metric to evolve smoothly away from critical points but to undergo a dramatic increase (or drop in other cases\cite{fidelityorigin2}) at them. By approaching a quantum critical point from either direction, the fidelity metric can be a powerful tool to signal a QPT/QCC.

In our numerical calculations, we find that the calculated fidelity metric is independent of the value of $\delta \lambda$ when $\delta \lambda$ is between $O$($10^{-5}$) and $O$($10^{-3}$), which is consistent with its intensive behavior; however, the fidelity metric could behave substantially differently when $\delta \lambda$ is smaller and larger, due to numerical errors and lack of accuracy, respectively. Thus, in what follows, we use $\delta\lambda=3\times10^{-4}$ to balance these two concerns.

\subsection{$d$-wave pair density matrix}

While the fidelity metric captures the existence of a QPT/QCC, we would also like to investigate quantities directly related to the order parameter of the ground-states. The $d$-wave pair density matrix~\cite{CNYang,Shastry1,LGor} can be defined as
\begin{equation}
P_{ij}^{d-wave}= \langle\Phi_0|D^{\dagger}_i D_j|\Phi_0\rangle
\end{equation}
One important quantity is the ratio between the largest and next largest eigenvalues of the $d$-wave pair density matrix, denoted as $R_{d-wave}=\Lambda_{1}(P_{ij}^{d-wave})/\Lambda_{2}(P_{ij}^{d-wave})$.
 The study of the ratio $R_{d-wave}$ was first suggested by RSH\cite{Shastry1,Shastry2}. To understand its significance, we note that based on the work of Penrose and Onsager~\cite{Penrose}, particle condensation into some well-defined ordered state leads to characteristic behavior in the corresponding density matrix: the largest eigenvalue scales to a finite value greater
than $ O(1)$ while the other eigenvalues are $\sim O(1)$. Further, Yang~\cite{CNYang}
 showed that the largest eigenvalue actually scales linearly with
 the total number of particles. Thus, $R_{d-wave}\sim\Psi^{2}N+\Phi$, where $\Psi$ is the order parameter and
   $\Phi$ is sub-linear in the system size.
   In other words, the ratio $R_{d-wave}$, having an intrinsic correlation to this
    $d$-wave SC order parameter, can be used directly to investigate $d$-wave
    SC order. A similar analysis follows for other (non-SC) orderings.

\subsection{Other quantities of interest}

Besides superconductivity, both charge and spin orders have been reported in cuprate high-temperature superconductors, with the undoped parent compounds known to be antiferromagnetic Mott insulators. Moreover, the 2D Hubbard model for sufficiently strong Coulomb repulsion $U$ possesses a well-known strong antiferromagnetic order at half-filling. Thus, we study charge/spin ordering  or the tendency toward ordering by calculating the charge and spin density matrices:
\begin{equation}\label{fidelity}
P_{ij}^{spin}= \langle\Phi_0|\sigma_{zi} \sigma_{zj}|\Phi_0\rangle,
\end{equation}
\begin{equation}\label{fidelity}
P_{ij}^{charge}= \langle\Phi_0|\rho_i \rho_j|\Phi_0\rangle,
\end{equation}
where $\sigma_{zi}=c_{i\uparrow}^{\dagger}c_{i\uparrow}-c_{i\downarrow}^{\dagger}c_{i\downarrow}$ and
$\rho_i=c_{i\uparrow}^{\dagger}c_{i\uparrow}+c_{i\downarrow}^{\dagger}c_{i\downarrow}$.
We investigate both $R_{spin}=\Lambda_{1}(P_{ij}^{spin})/\Lambda_{2}(P_{ij}^{spin})$ and $R_{charge}=\Lambda_{1}(P_{ij}^{charge})/\Lambda_{2}(P_{ij}^{charge})$, as we do for the $d$-wave pair density matrix. We also examine the specific matrix elements of the density matrix associated with the correlation functions. By calculating the above quantities, we investigate the strongest order that exists in proximity to the $d$-wave SC order in the small cluster, to provide insights into the behavior of competing or coexisting states in the thermodynamic limit.

\section{Results}

Before describing our numerics, we clarify our viewpoint. Based on the experience in RSH, we look at the fidelity metric $g$, together with the ratio $R$ of the largest to next largest eigenvalues of the density matrix with suitable symmetry. It is a combined look at these variables $g$ and $R$ as functions of $\lambda$ that enables us to draw conclusions about the nature of ordering. We note that a peak in $g$ usually signifies a crossover between two phases, and a jump in $g$ signifies a level crossing. We denote by $\lambda^{*}$ the location of the peak in $g(\lambda)$, and it is seen below to play an important role in the analysis. The ratio $R$ has a magnitude $\sim 1$, when the symmetry of the order parameter is incompatible with the ground-state. It is large (3-10 in the $t$-$J$ model studied in RSH) when the symmetry is compatible, for small systems. In the limit of large systems,
 which is currently unattainable, it is not possible to assert that a quantum critical point exists,
  but looking at $g$ and $R$ together gives us a reasonable picture of the evolution
   of the ground-state.

We perform the calculations with the ED technique in a finite-size cluster using the Parallel ARnoldi PACKage (PARPACK)\cite{PARPACK} based on the Implicitly Restarted Arnoldi method. This approach is efficient in obtaining a set of properly orthonormalized eigenvectors and is more numerically stable compared with standard Lanczos methods. In our ED problem, the size of the Hilbert space grows exponentially with the increase of the cluster size. Due to computational time considerations, we perform our calculations for the 16B Betts cluster\cite{BettsCluster} with periodic boundary conditions. Moreover, the 16B Betts cluster provides access to momentum ($\pi/2,\pi/2$), which is important in catching low energy excitations at half-filling and low dopings. Considering symmetries, including the invariance of the total number of electrons and total momentum, and fixing ourselves in the space with total spin $S_z = 0$, the Hilbert space size
of the Hamiltonian is diminished to $\sim 10^7$ so that the ground-state eigenvector and eigenvalue could be calculated by ED using PARPACK in a reasonable time.

\subsection{Antiferromagnetic Order at Half-filling}

\begin{figure}
\includegraphics[width=0.9\columnwidth]{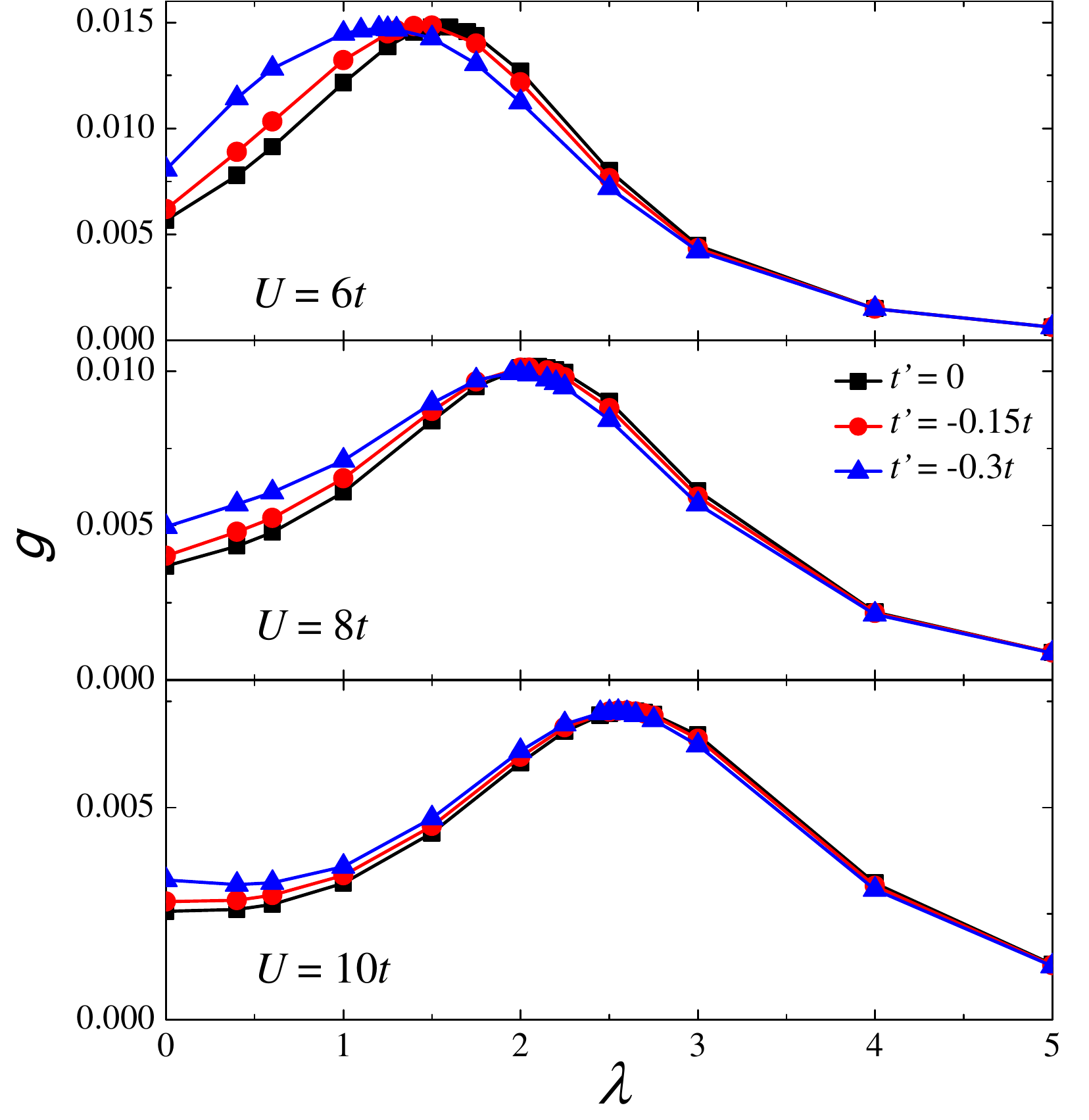}
\caption{\label{Fig:1} The fidelity metric $g$ as a function of $d$-wave
pair field strength $\lambda$ (in units of $t$) at half-filling. Coulomb $U(6t,8t,10t)$ and $t'(0,-0.15t,-0.3t)$ are indicated. The small magnitude of $g$ indicates the incompatibility of $d$-wave order at half-filling.}
\end{figure}
The fidelity metric $g$ at half-filling for various values of $U$ and $t^\prime$ (measured in units of t) is shown in Fig.~1. $g$ approaches a small value as $\lambda\rightarrow0$, and $g\rightarrow0$ as $\lambda$ increases beyond the peak value. These two regions correspond to two different phases: the $d$-wave SC state at large $\lambda$ and the competing state at small $\lambda$. $\lambda$* is larger than $1$, implying that this competing state exhibits strong order which is even stronger as $U$ increases, since it requires a large $d$-wave $\lambda$ to conquer the competing phase.

To obtain a full picture, we also calculate both $R_{d-wave}$ and $R_{spin}$ as functions of $\lambda$ with the results shown in Fig.~2. First, $R_{d-wave}$ reaches $1$ as $\lambda\rightarrow 0$, which confirms the incompatibility of $d$-wave SC order in the Hubbard model at half-filling. Second, the large value of $R_{spin}$ at $\lambda\rightarrow0$ is consistent with the well known fact that the 2D Hubbard model at half-filling exhibits strong antiferromagnetic order. The values of $R_{spin}$ at $\lambda\rightarrow0$ increase for increasing $U$, demonstrating stronger antiferromagnetic order for larger values of $U$. Further examining the spin correlation functions (not shown) reveals a $(\pi,\pi)$ antiferromagnetic order. Third, with the increase in $\lambda$, $R_{spin}$ decreases and $R_{d-wave}$ increases, with a crossover at a certain $\lambda$. The crossover of $R_{spin}$ and $R_{d-wave}$ in Fig.~2 moves to higher $\lambda$ at larger $U$, consistent with the $\lambda$* from the fidelity metric $g$ in Fig.~1. Based on these points, the fidelity $g$ in Fig.~1 is now more clearly understandable. The small peak at $\lambda$* between 1 to 2.5 for various parameters arises from forcing $d$-wave SC order by the applied SC term in Eq.~(2). The peak should grow with increasing cluster size to indicate a true QPT, and thus for our cluster the peak implies a crossover (QCC) rather than a true QPT. In a strict sense we are thus studying QCC rather than quantum critical points.

We next discuss the effects of $t^\prime$. Figure 1 shows that the QCC moves to lower values of $\lambda$ as the magnitude of $t^\prime$ increases. Moreover, Fig.~2 demonstrates that negative $t^\prime$ enhances only slightly the $d$-wave SC order measured by $R_{d-wave}$ while substantially suppressing the antiferromagnetic order as reflected in $R_{spin}$. This indicates that while negative $t^\prime$ may hurt the antiferromagnetic order, it does not necessarily lead to an enhancement of superconductivity at half filling. These effects in $R_{spin}$ and $R_{d-wave}$ do not depend critically on the value of $U$ for this range of parameters.
\begin{figure}
\includegraphics[width=0.95\columnwidth]{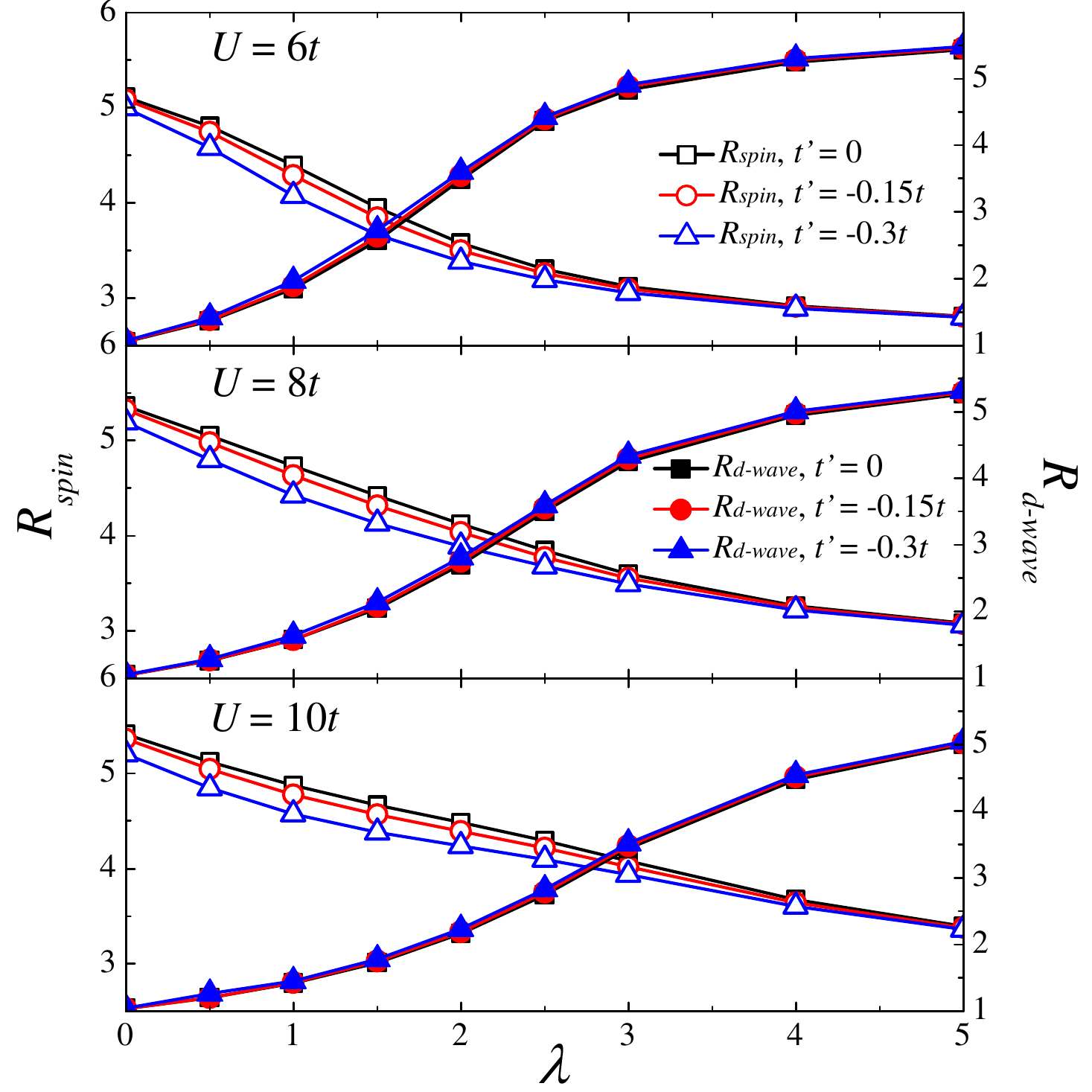}
\caption{\label{Fig:2} The ratio of the largest to second largest eigenvalue of the $d$-wave pair density matrix $R_{d-wave}$ and that of the spin density matrix $R_{spin}$ as functions of $d$-wave pair field strength $\lambda$ (in units of $t$) at half-filling. Coulomb $U(6t,8t,10t)$ and $t'(0,-0.15t,-0.3t)$ are indicated.}
\end{figure}

\subsection{Hole-doping near Half-filling}

Figure 3 shows results at 12.5\% hole-doping that follow a similar trend as half-filling: as $\lambda\rightarrow0$, $g$ approaches a small finite value; as $\lambda$ increases, $g$ increases to a maximum at some $\lambda$* and then decreases monotonically toward zero with additional increase in $\lambda$. This behavior indicates that the ground-state of the Hubbard model is not a $d$-wave SC for this cluster size or parameter regime. The dependence of $\lambda$* on $U$ and $t^\prime$ is also similar to the results at half-filling, except for the discontinuities in $g$ for $t^\prime=-0.3t$ at $\lambda\sim1.0$, stemming from an abrupt change in the momentum sector of the ground-state. This type of sudden change for these sets of parameters only occurs for this specific cluster and should not occur in the thermodynamic limit. The trends in $R_{d-wave}$ and $R_{spin}$ as functions of $\lambda$ at various values of $U$ and $t^\prime$ are also similar to the results at half-filling shown in Fig.~2 --- $d$-wave SC and spin order depend mainly on $U$ and slightly on $t^\prime$. As seen in Fig.~4, $R_{spin}$ is larger than 1 as $\lambda\rightarrow0$; however, it is smaller here than at half-filling, implying that the competing state is also antiferromagnetic but weaker than that at half-filling. The sudden jump in $R_{spin}$ for $t'=-0.3t$ in Fig.~4 occurred for the same reason as that in $g$, as we have discussed.

In contrast to half-filling, at this doping negative $t^\prime$ suppresses antiferromagnetism and slightly enhances superconductivity, as shown in the larger value of $R_{d-wave}$ and the smaller value of $R_{spin}$ for negative $t^\prime$ at $\lambda\sim 0-2$ in Fig.~(4). These trends exist for all the values of $U$ in our calculations.

\begin{figure}
\includegraphics[width=0.9\columnwidth]{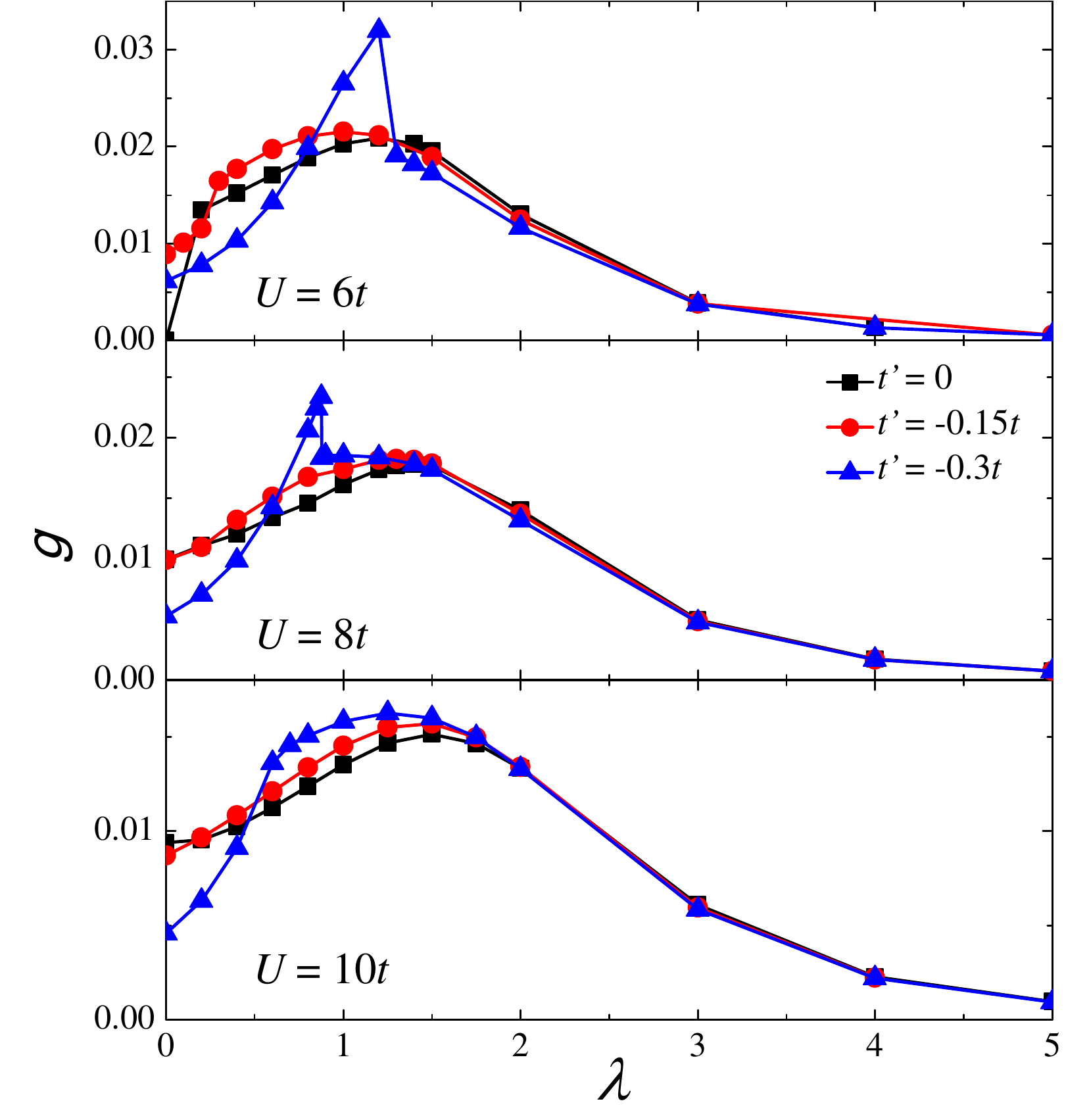}
\caption{\label{Fig:3} The fidelity metric $g$ as a function of $d$-wave
pair field strength $\lambda$ (in units of $t$) at $12.5\%$ hole-doping. Coulomb $U(6t,8t,10t)$ and $t^\prime(0,-0.15t,-0.3t)$ are indicated.}
\end{figure}
\begin{figure}
\includegraphics[width=0.95\columnwidth]{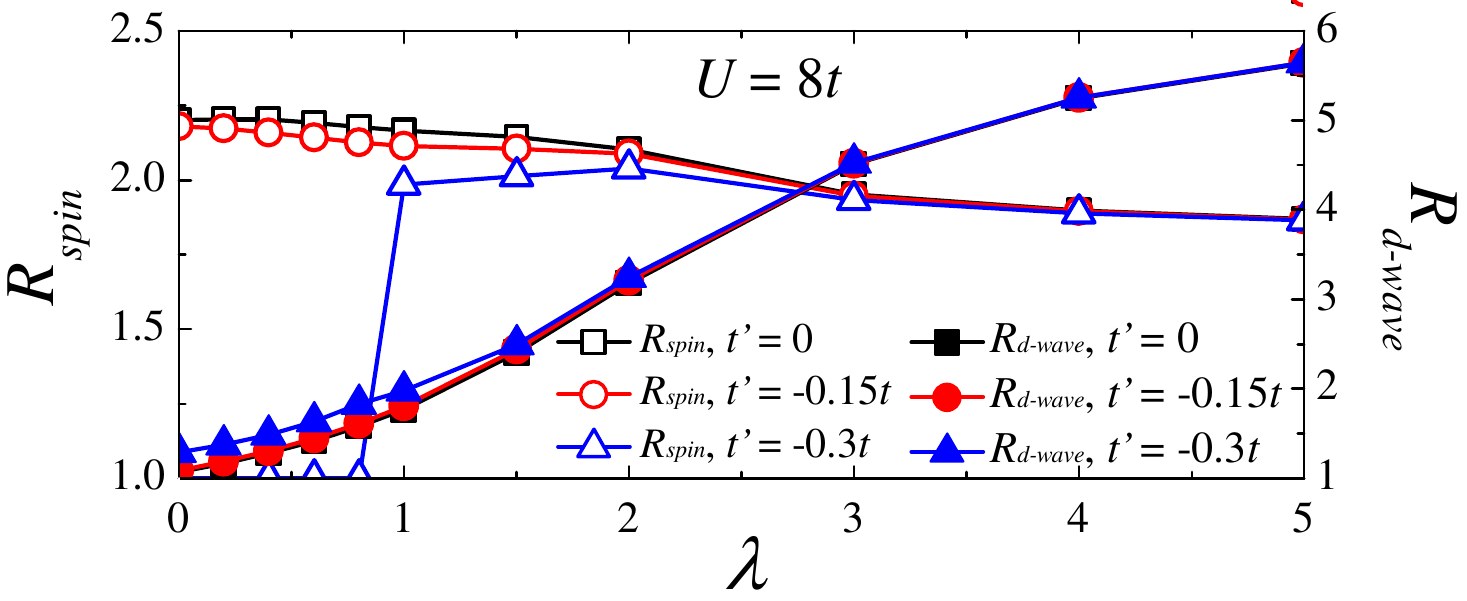}
\caption{\label{Fig:4} The ratios of the largest to second largest eigenvalue of the $d$-wave pair density matrix $R_{d-wave}$ and that of the spin density matrix $R_{spin}$ as functions of $d$-wave pair field strength $\lambda$ (in units of $t$) at $12.5\%$ hole-doping. Coulomb $U=8t$ and $t^\prime(0,-0.15t,-0.3t)$ are indicated.}
\end{figure}

\subsection{SC state at $25\%$ hole-doping}

Figure 5 shows the fidelity metric calculations at 25\% hole-doping, with a log-linear scale employed to catch clearly the transition as $\lambda\rightarrow 0$. As seen in Fig.~5, $\lambda$* moves close to zero or even negative values (see also Fig.~13), suggesting that the state remains in the same phase as $\lambda$ decreases toward zero. In Fig.~6, the robust value of $R_{d-wave}$$\sim$ 10 for $\lambda\sim 3$ is an indicator of $d$-wave SC order; $R_{d-wave}$ persists to $\sim$ 2 (but is not as small as $\sim$ 1) for $\lambda \sim 0$. This strongly suggests the existence of $d$-wave SC order in the 2D single-band Hubbard model even without the $d$-wave pair field. We note that this is supported by recent calculations based on infinite projected entangled-pair states in the $t$-$J$ model which yield the strongest SC pairing around 25\% doping. \cite{tJEntangledPairStates}

There is growing evidence that $t^\prime$ plays an important role in the critical temperature and other properties of SC materials\cite{longhopping1,Tanakat',Ref9Aadded,Ref9Badded, t'hurts,t'hurtsWhite,ThMaiert',ttJpair,t'hurtsGFMC,t'HubbardHolstein,DCAKentt',t'Sakakibara}. Our calculations support the statement that negative $t^\prime$ favors a $d$-wave SC state, indicated by the fidelity metric calculations shown in Fig.~5, in which negative $t^\prime$ gives a rapid increase to the fidelity metric as $\lambda$ goes to zero and yields $\lambda$* $\rightarrow$ 0 with increasing magnitude of $t^\prime$, and also demonstrated by the $d$-wave pair density matrix calculations, in which $R_{d-wave}$ (at $\lambda=0$) has a larger value as the magnitude of $t^\prime$ increases. Notice there are sudden changes of $g$ at $t^\prime=0 and U=6t,8t$ or $10t$, only near $\lambda\rightarrow0$, which do not originate from a change of the momentum sector of the ground-state as encountered in previous cases. These sudden changes are seen in Fig.~6 for $R_{d-wave}$ as well, where $R_{d-wave}$ jumps to $1$ as $\lambda\rightarrow0$, where the eigenstates of the $d$-wave pair density matrix with the largest eigenvalue are two-fold degenerate.


\begin{figure}
\includegraphics[width=0.9\columnwidth]{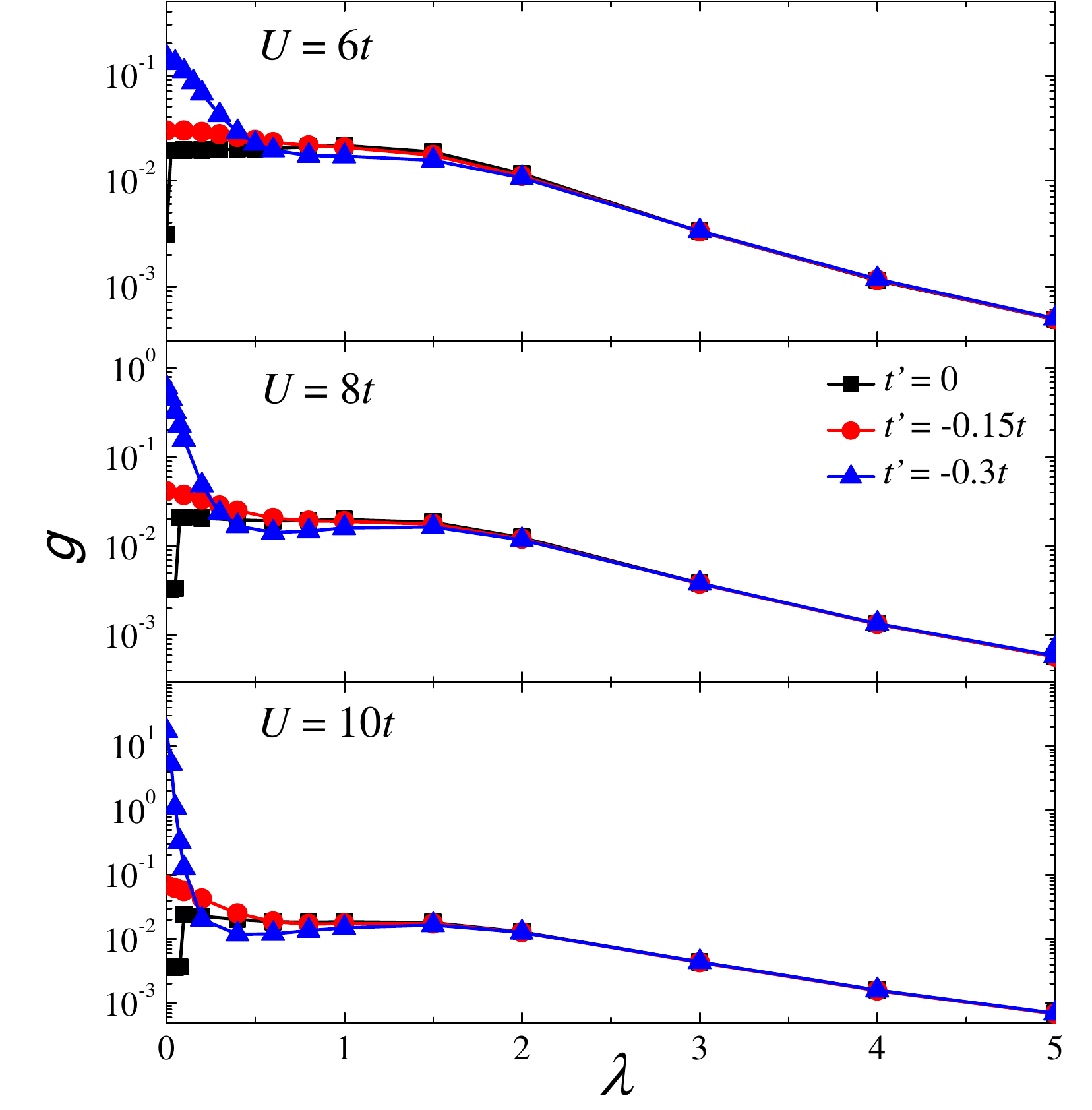}
\caption{\label{Fig:5} The fidelity metric $g$ as a function of $d$-wave
pair field strength $\lambda$ (in units of $t$) at $25\%$ hole-doping. Coulomb $U(6t,8t,10t)$ and $t^\prime(0,-0.15t,-0.3t)$ are indicated. A semi-log scale is employed to highlight the behaviors around $\lambda=0$.}
\end{figure}
\begin{figure}
\includegraphics[width=0.95\columnwidth]{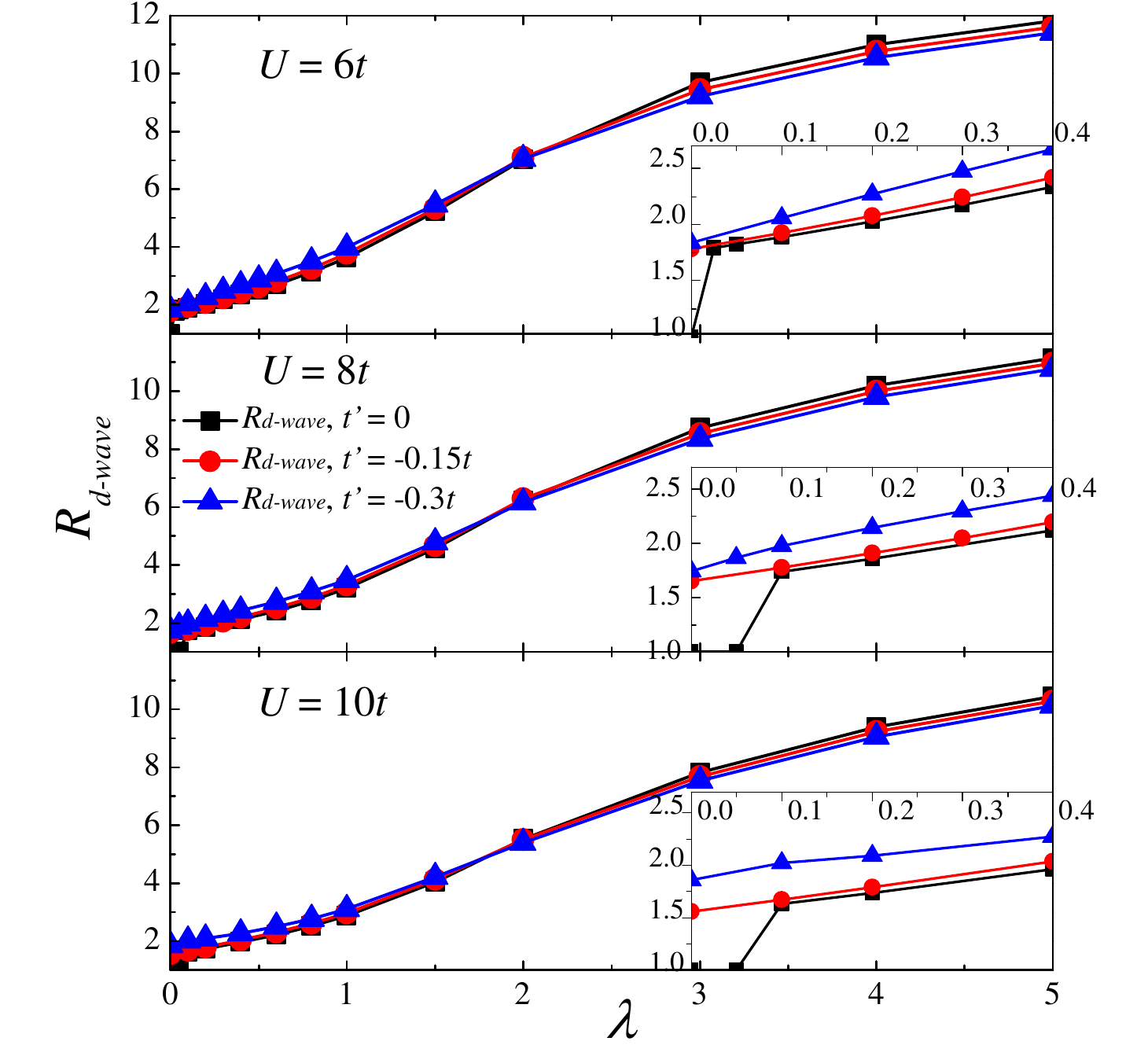}
\caption{\label{Fig:6} The ratio of the largest to second largest eigenvalue of the $d$-wave pair density matrix $R_{d-wave}$ as a function of $d$-wave pair field strength $\lambda$ (in units of $t$) at 25\% hole-doping. Coulomb $U(6t,8t,10t)$ and $t^\prime(0,-0.15t,-0.3t)$ are indicated. The behaviors near $\lambda=0$ are highlighted in the insets.}
\end{figure}

\subsection{Hole-doping far from Half-filling}

In the highly doped region ($37.5$ and $50\%$ hole-doping), fidelity metric calculations display positive $\lambda$*, indicating a ground-state incompatibility with $d$-wave pairing as $\lambda\rightarrow0$, and give a $\lambda$* that depends strongly on $t^\prime$: at $37.5\%$ hole-doping [Fig.~7], $\lambda$*$\sim0.25, 0.5$ and $0.75$ at $t^\prime=-0.3t, -0.15t$ and $0$; at $50\%$ hole-doping [Fig.~8], $\lambda$*$\sim0.25, 0.4$ and $0.6$ at $t^\prime=-0.3t, -0.15t$ and $0$. Since $\lambda$* here is one order of magnitude smaller than the $\lambda$* in the weakly doped region, the competing state here is more easily suppressed by the $d$-wave pair field when compared with the strong antiferromagnetic order exhibited near half-filling.
\begin{figure}
\includegraphics[width=0.9\columnwidth]{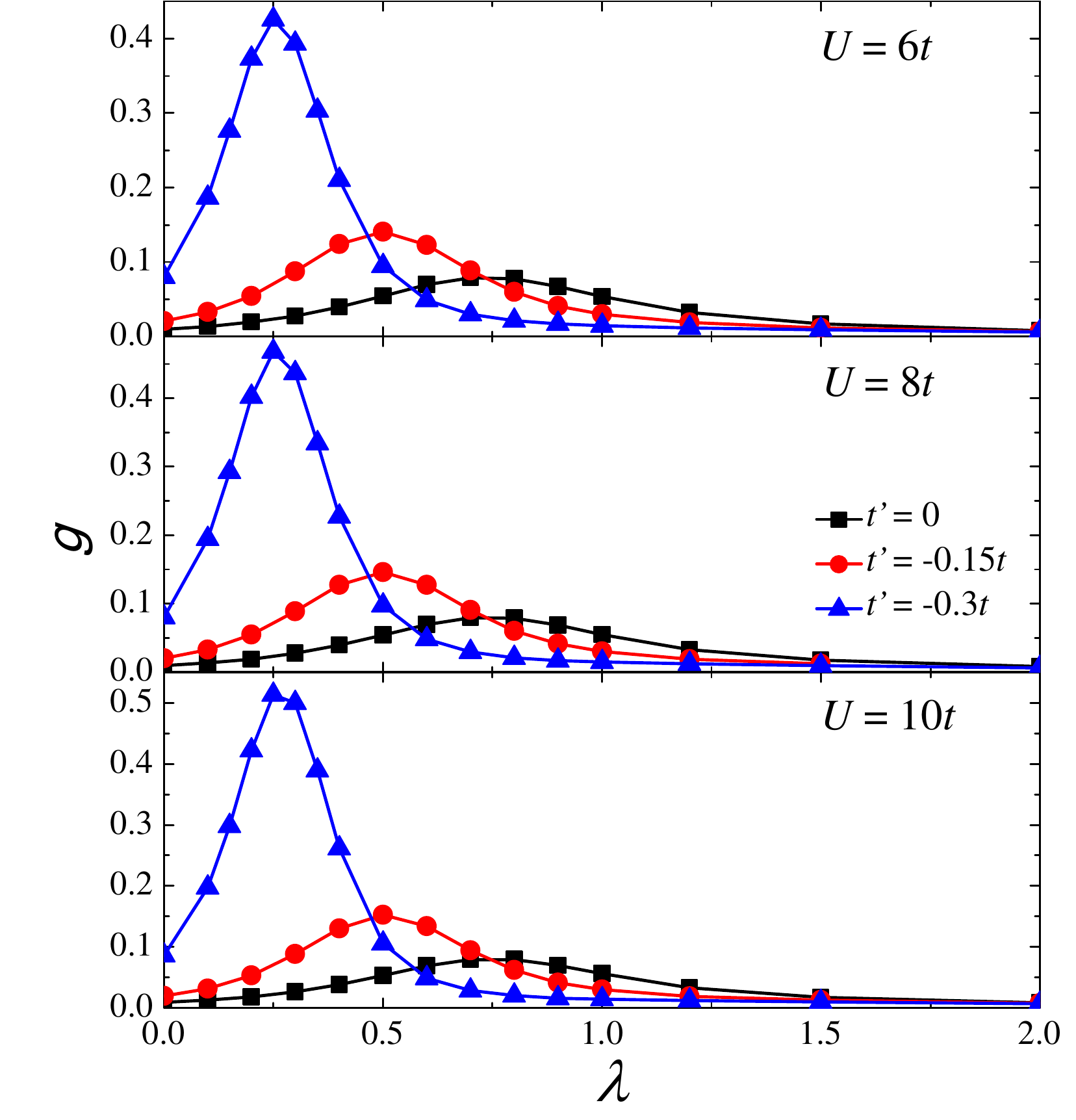}
\caption{\label{Fig:7} The fidelity metric $g$ as a function of $d$-wave
pair field strength $\lambda$ (in units of $t$) at $37.5\%$ hole-doping. Coulomb $U(6t,8t,10t)$ and $t^\prime(0,-0.15t,-0.3t)$ are indicated.}
\end{figure}
\begin{figure}
\includegraphics[width=0.9\columnwidth]{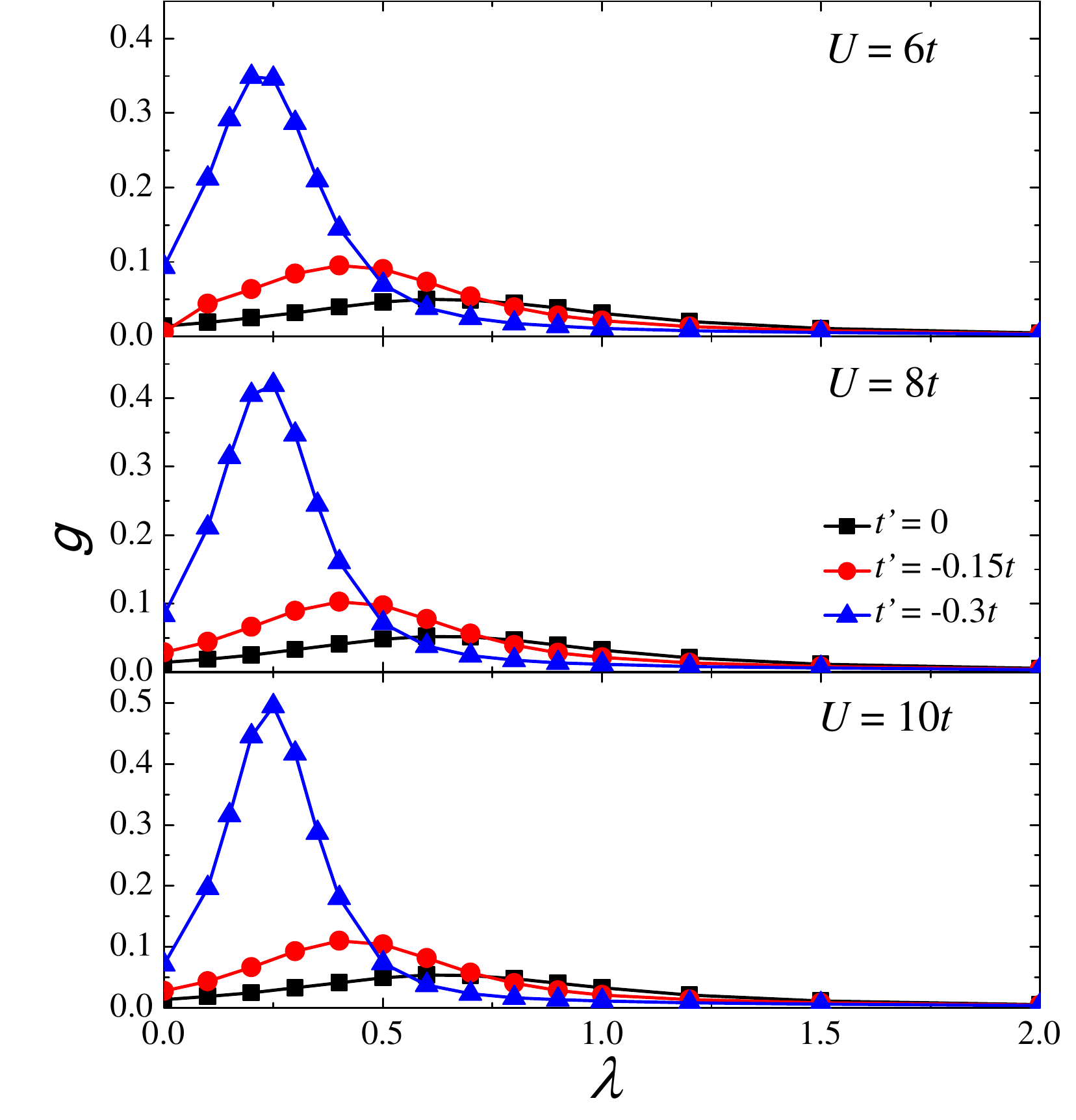}
\caption{\label{Fig:8} The fidelity metric $g$ as a function of $d$-wave
pair field strength $\lambda$ (in units of $t$) at $50\%$ hole-doping. Coulomb $U(6t,8t,10t)$ and $t^\prime(0,-0.15t,-0.3t)$ are indicated.}
\end{figure}
\begin{figure}
\includegraphics[width=\columnwidth]{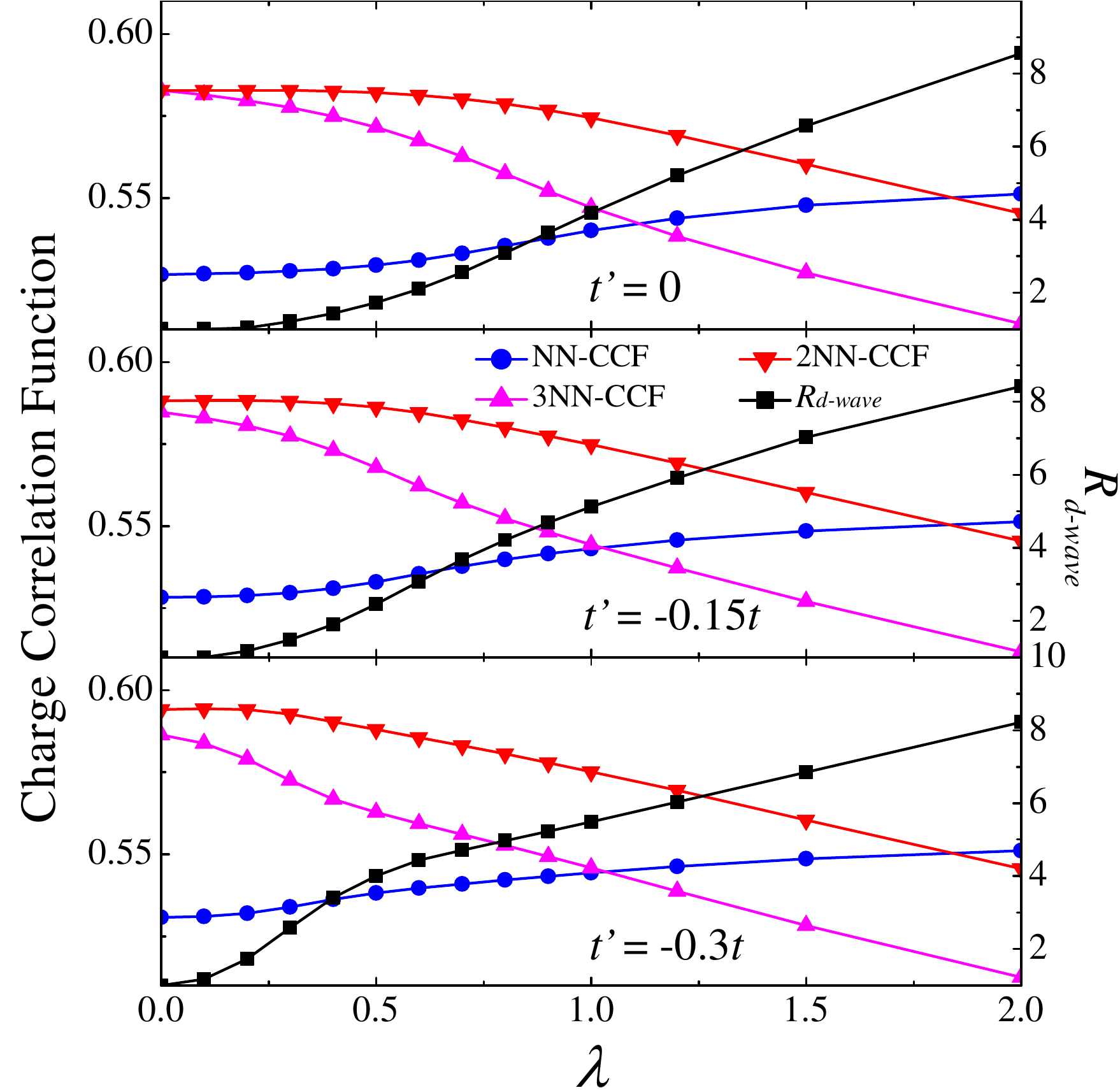}
\caption{\label{Fig:9} The CCFs (NNCCF, 2NNCCF and 3NNCCF) and the ratio of the largest to second largest eigenvalue of the $d$-wave pair density matrix $R_{d-wave}$ as functions of $d$-wave pair field strength $\lambda$ (in units of $t$) at 37.5\% hole-doping. Coulomb $U=10t$ and $t^\prime(0,-0.15t,-0.3t)$ are indicated.}
\end{figure}
\begin{figure}
\includegraphics[width=\columnwidth]{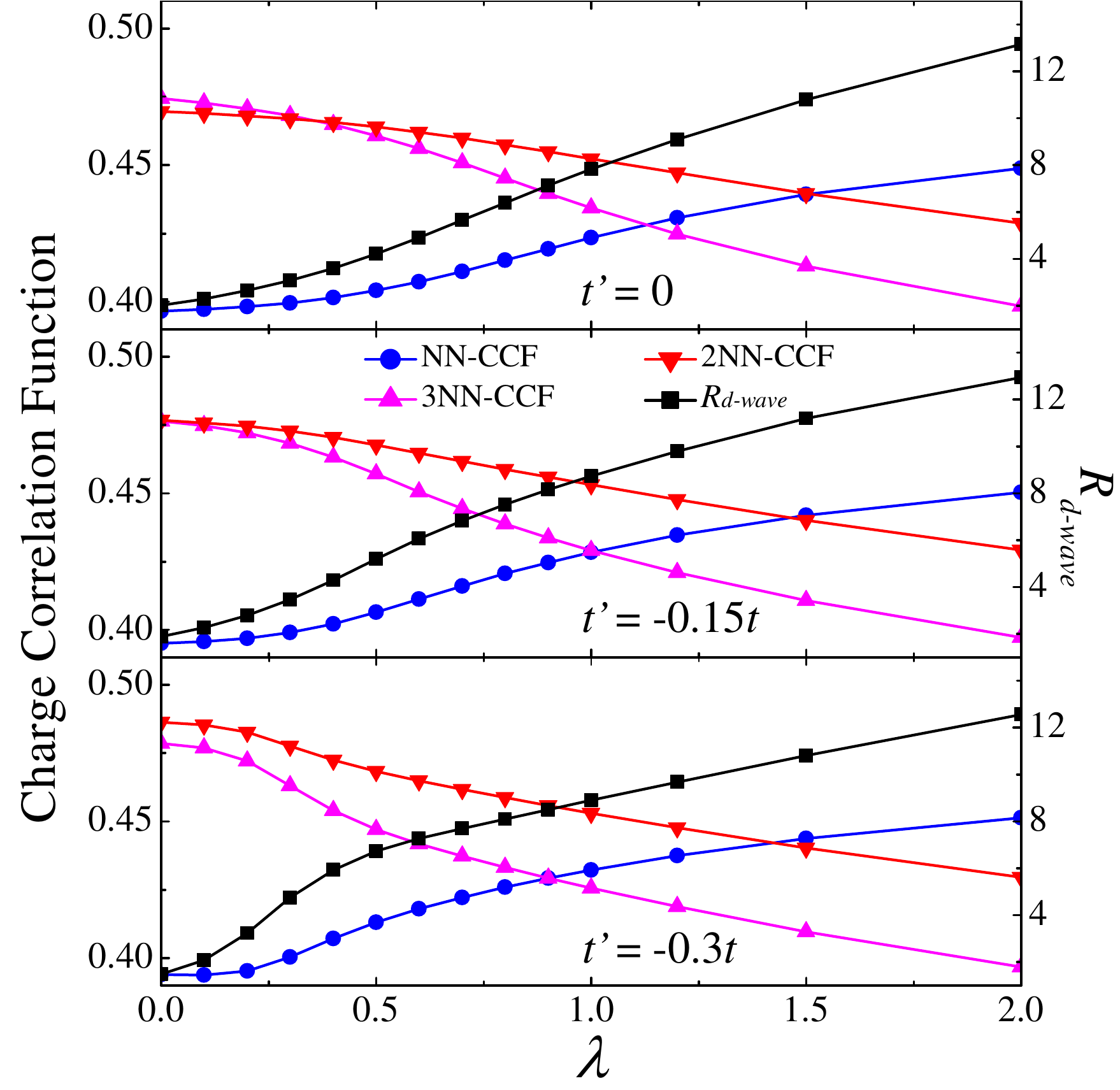}
\caption{\label{Fig:10} The CCFs (NNCCF, 2NNCCF and 3NNCCF) and the ratio of the largest to second largest eigenvalue of the $d$-wave pair density matrix $R_{d-wave}$ as functions of $d$-wave pair field strength $\lambda$ (in units of $t$) at 50\% hole-doping. Coulomb $U=10t$ and $t^\prime(0,-0.15t,-0.3t)$ are indicated.}
\end{figure}
\begin{figure}
\includegraphics[width=\columnwidth]{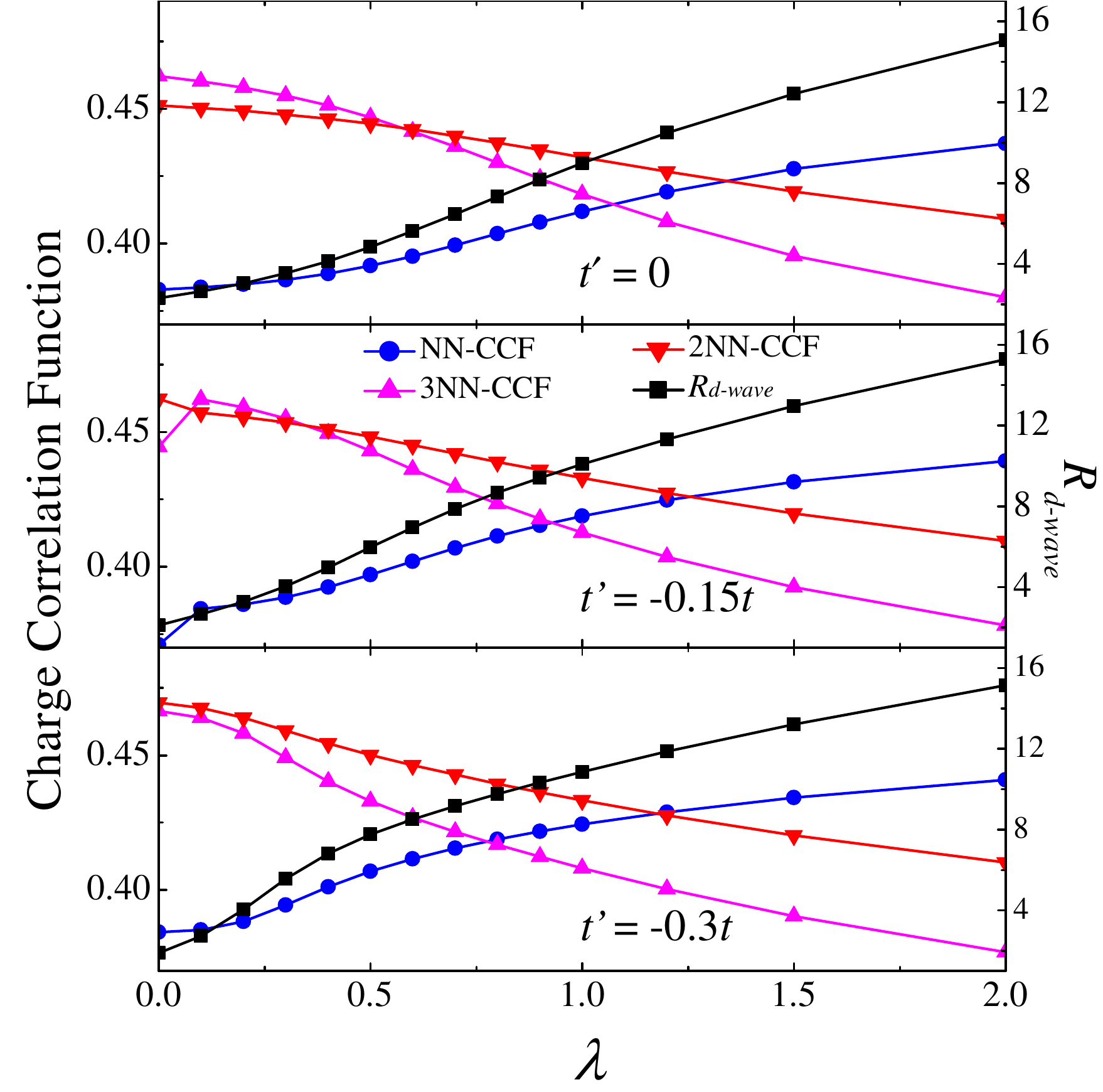}
\caption{\label{Fig:11} The CCFs (NNCCF, 2NNCCF and 3NNCCF) and the ratio of the largest to second largest eigenvalue of the $d$-wave pair density matrix $R_{d-wave}$ as functions of $d$-wave pair field strength $\lambda$ (in units of $t$) at 50\% hole-doping. Coulomb $U=6t$ and $t^\prime(0,-0.15t,-0.3t)$ are indicated.}
\end{figure}

Calculations of the charge density matrix show that $R_{charge}$ is always greater than $1$ and does not change much as a function of $\lambda$, so tracking $R_{charge}$ by itself might be insufficient to determine the change of charge order. We then focus on the charge correlation functions (CCFs) --- the elements in the charge density matrix $P_{ij}^{charge}=\langle\Phi_0|\rho_i \rho_j|\Phi_0\rangle$ after being normalized. We investigate the normalized nearest-neighbor CCF (NNCCF) $\langle\Phi_0|\rho_r \rho_{r+\hat{x}}|\Phi_0\rangle/\langle\Phi_0|\rho_r \rho_r|\Phi_0\rangle$, the second-NNCCF 2NNCCF $\langle\Phi_0|\rho_r \rho_{r+\hat{x}+\hat{y}}|\Phi_0\rangle/\langle\Phi_0|\rho_r \rho_r|\Phi_0\rangle$ and the third-NNCCF 3NNCCF $\langle\Phi_0|\rho_r \rho_{r+2\hat{x}}|\Phi_0\rangle/\langle\Phi_0|\rho_r \rho_r|\Phi_0\rangle$, and we observe how they evolve as functions of $\lambda$. Figures 9 through 11 show the normalized CCFs and $R_{d-wave}$ at 37.5\% doping for $U=10t$, 50\% doping for $U=10t$ and 50\% doping for $U=6t$. We observe a QCC between the $d$-wave SC state and some charge ordered state by the trends in $R_{d-wave}$ and the CCFs as $\lambda$ decreases toward $0$ ($R_{d-wave}$ and NNCCF drop, and the other CCFs increase). This is a QCC from the $d$-wave SC state to a checkerboard charge ordered state. Careful examinations of CCFs at $\lambda=0$ show that this charge order for the bare Hubbard model is short-ranged or local, since CCFs at longer distances (up to the half of the diagonal of this cluster - the largest possible distance for CCFs) do not exhibit distinctive charge order (not shown). Therefore, our calculation shows a short-range checkerboard charge ordered state at large hole-doping in the 2D Hubbard model. The evolution from the local checkerboard charge order to the $d$-wave SC order, as $\lambda$ increases, originates from the fact that Cooper pair formation favors the nearest-neighbor pairing or charge correlation, which destroys the checkerboard charge order where the nearest-neighbor charge correlation is small. The transition at $50\%$ doping (see Fig.~10) is similar to that at $37.5\%$ doping. One interesting finding is that this QCC is dependent not on Coulomb $U$ but rather on $t^\prime$, demonstrated by the comparison between the calculations for different values of $U$, as shown in Fig.~10 and Fig.~11. At this doping, electron density is low and double occupancy is less relevant, so Coulomb $U$ has little effect on the properties of the ground-state.

Our results show that as we turn on $\lambda$ and observe how the local charge ordered state competes with $d$-wave superconductivity, negative $t^\prime$ seems to only slightly suppress the charge order but rather largely reinforce the superconductivity. This is indicated by comparing $R_{d-wave}$ at different $t^\prime$, where $R_{d-wave}$ at a large magnitude of $t^\prime$ grows much faster than at a small magnitude of $t^\prime$ as $\lambda$ is turned on.


\begin{figure}
\includegraphics[width=0.9\columnwidth]{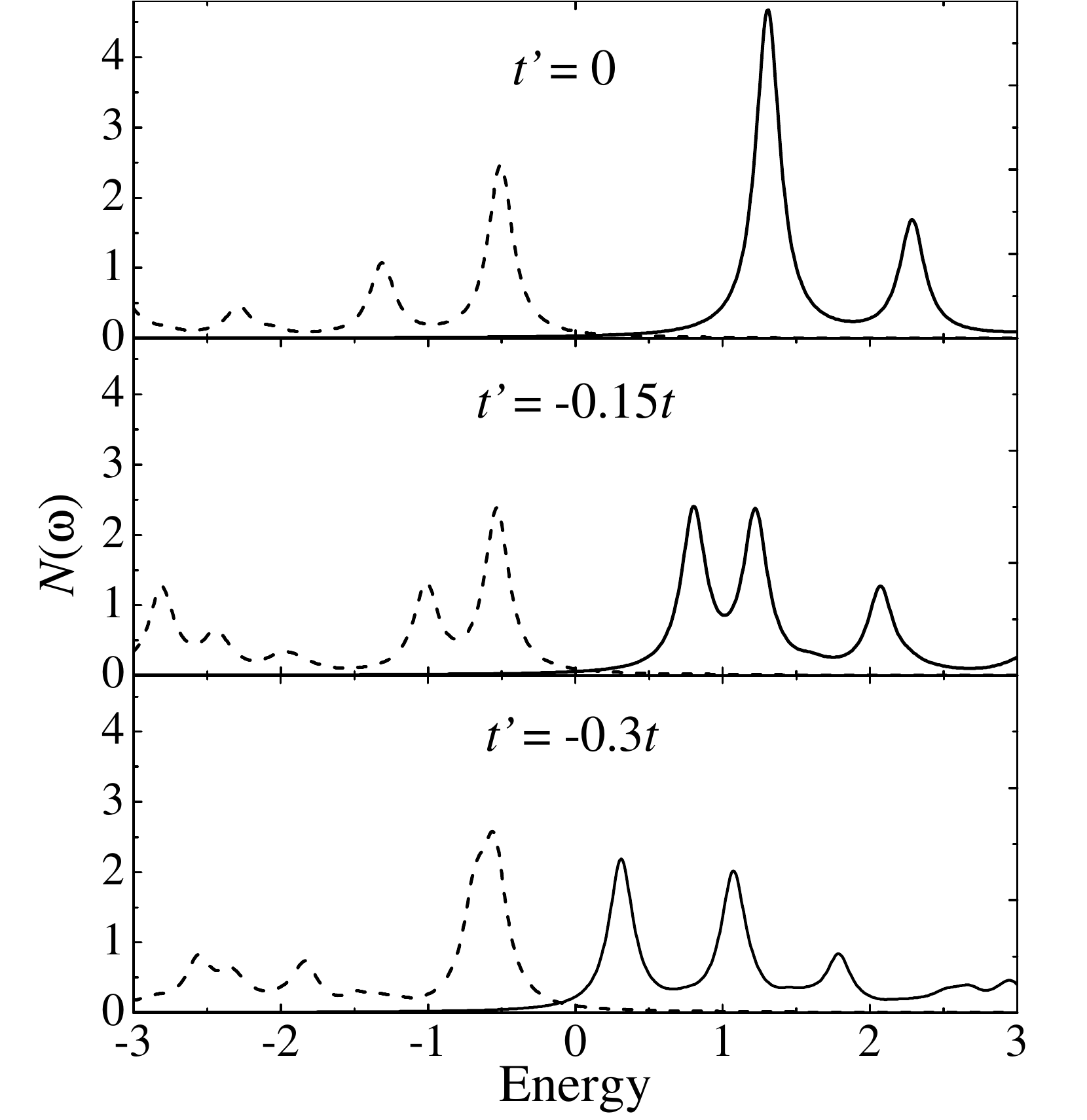}
\caption{\label{Fig:12} The density of states for 16-site single-band Hubbard model without d-wave pair field at 37.5\% hole-doping at $U=8t$ and $t^\prime(0,-0.15t,-0.3t)$, energy is in units of $t$. Energy zero represents the ground-state energy. Dash line and solid line represent the occupied and unoccupied states respectively.}
\end{figure}

The way negative $t^\prime$ enforces the $d$-wave superconductivity might be related to the suppression of antiferromagnetic order near half-filling, as discussed in Sec. III.A and III.B. We also note that the gathering of density of states in proximity of the Fermi level at negative $t^\prime$ for our parameter range may also help form SC order. To justify this point, we calculate the density of states for the pure Hubbard model at $37.5\%$ hole-doping, as shown in Fig.~12. This is done by summing over the spectral functions at all momentum points from our 16-site calculations. The dashed lines show the removal spectra or the occupied states, and the solid lines show the addition spectra or the unoccupied states. A more negative $t^\prime$ facilitates a gathering of density of states in both the removal and addition spectra at the proximity of the Fermi level (which can be defined as the crossover of the dashed line and the solid line). The gathering of density of states near the Fermi level helps formation of superconductivity by providing more weight to mix particles and holes to form the condensate.\cite{BCS} Particularly at high hole-doping, this trend prevails over any effect that $t^\prime$ may have on antiferromagnetic ordering or charge density wave tendencies. As the density of states at the Fermi level is proportional to the SC susceptibility, our calculations indicate a stronger susceptibility for negative $t^\prime$ in the high hole-doping regime for our parameter range.

\section{Phase Diagram, Summary and Conclusion}

Based on the above calculations at various dopings, we obtain the Hubbard model phase diagram tracking $\lambda$* as a function of doping, as shown in Fig.~13. Combined with the $d$-wave pair, charge, and spin density matrix calculations, several conclusions are obtained:
(i) At half-filling and low hole-doping, the Hubbard model exhibits a strong antiferromagnetic order with $\lambda$* depending primarily on Coulomb $U$; (ii) for high hole-doping, we observe short-range checkerboard charge order with $\lambda$* exclusively determined by $t^\prime$; (iii) in our cluster at 25\% hole-doping, $\lambda$* moves close to zero or even negative values, demonstrating a $d$-wave SC state in the pure Hubbard model in this cluster, supported by other numerical studies\cite{tJEntangledPairStates}; (iv) negative $t^\prime$ favors $d$-wave SC order, where $\lambda$* is smaller at larger absolute value of $t^\prime$ and where $R_{d-wave}$ has a larger value at $\lambda=0$ for larger absolute value of $t^\prime$; and (v) the antiferromagnetism at half-filling and low doping is much stronger than the short-range checkerboard charge order for high hole-dopings, revealed by a much larger $\lambda$* in the former case.

The phase diagram has several compelling features that indicate that the Hubbard model contains rich physics that may underlie an understanding of different competing phases in the cuprates. The various phases that our calculations suggest follow trends similar to those from recent density matrix renormalization group calculations on the infinite $U$ Hubbard model\cite{LiuInfiniteU} and infinite projected entangled-pair states calculations in the $t$-$J$ model\cite{tJEntangledPairStates}. Clearly, our results show the prevalence of antiferromagnetic order around half-filling and that negative $t^\prime$ only slightly assists superconductivity while largely suppressing antiferromagnetism. Here, the dominant energy scale is the Hubbard $U$. At large doping, a short-range checkerboard charge order emerges. This short-range checkerboard charge order is more easily destabilized by the $d$-wave pair field than antiferromagnetism, giving smaller values of $\lambda$* at large hole-dopings. Here, negative $t^\prime$ largely facilitates superconductivity and slightly suppresses checkerboard charge order. In between, it appears that superconductivity may emerge as a consequence of the battle between antiferromagnetism controlled by $U$ and density of states and possibly checkerboard charge order controlled by $t^\prime$. Therefore, it is tempting to intuit that the physics of $t^\prime$ may not be reflective of orbital content (such as apical oxygen content) or a destabilization of the Zhang-Rice singlet but more connected to a destabilization of a competing spin/charge order and density of states effects at low and high hole-doping and a reinforcement of the SC order at low and high hole-doping. Thus, it would be of great interest to explore larger cluster sizes to examine the changes of the phase diagram as a function of cluster size.

\begin{figure}
\includegraphics[width=\columnwidth]{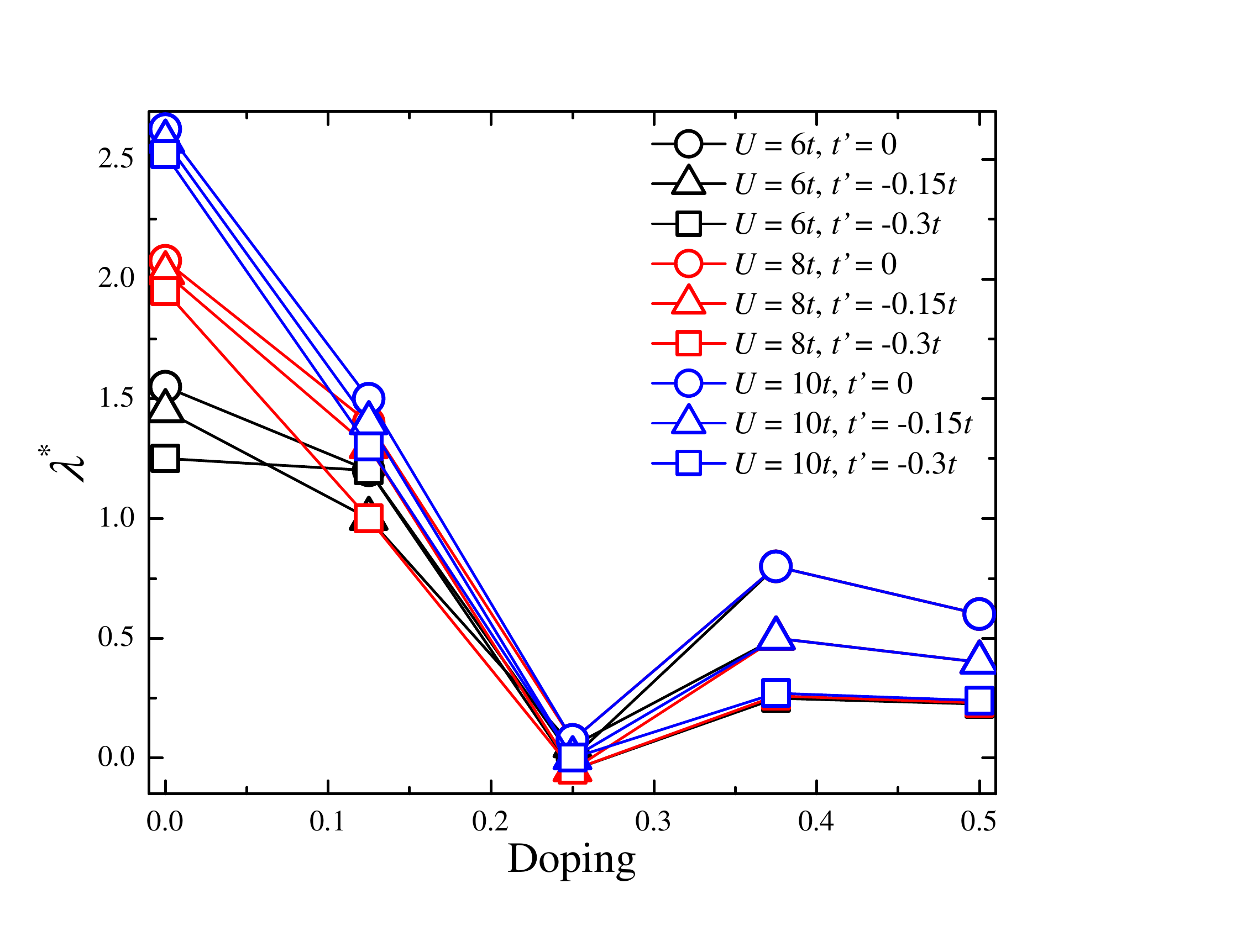}
\caption{\label{Fig:13} Phase diagram of $\lambda$* (in units of $t$) vs.~hole-doping rate at $U=6t,8t,10t$ and $t^\prime=0,-0.15t,-0.3t$. The 2D single-band Hubbard model favors a $d$-wave SC phase at $25\%$ hole-doping, an antiferromagnetic phase at low doping and local checkerboard charge order at high doping. Strong antiferromagnetic order near half-filling is mainly dependent on Coulomb $U$, while the strength of the local checkerboard charge order far from half-filling is exclusively determined by $t^\prime$.}
\end{figure}



In summary, ED techniques have been employed to
study the $d$-wave SC phase of the 2D single-band Hubbard model.
The fidelity metric $g$, the $d$-wave pair and charge/spin density matrices, including $R_{d-wave}$, $R_{spin}$, and CCFs,
have been calculated for the single-band Hubbard model plus
a $d$-wave pair field term at various hole-dopings and
parameters, including Coulomb $U$, the next-nearest-neighbor
hopping $t^\prime$, and the $d$-wave SC pair field strength $\lambda$.
Calculations show that at 25\% hole-doping the 2D single-band Hubbard model possesses a $d$-wave SC ground-state. Based on these results, we further summarize the behavior of $\lambda$* as a function of doping in the phase diagram. Moreover, the phase diagram shows that $d$-wave SC order is enhanced by negative $t^\prime$, which is consistent with
experiments\cite{longhopping1,Tanakat',Ref9Aadded,Ref9Badded}.
Finally, to obtain the $d$-wave SC order parameter, finite-size
scaling analysis could be done, provided that ED calculations
on larger systems become feasible in the future.

\section{Acknowledgements}

We would like to acknowledge S.~A.~Kivelson and L.-Q.~Lee for valuable discussions. This work was supported at SLAC National Accelerator Laboratory and Stanford University by the U.S. Department of Energy, Office of Basic Energy Sciences, under Contract No.~DE-AC02-76SF00515 and at University of California, Santa Cruz under Contract No.~DE-FG02-06ER46319. C.~J.~Jia is also supported by the Stanford Graduate Fellows in Science and Engineering Program. A portion of the computational work was performed using the resources of the National Energy Research Scientific Computing Center, supported by the U.S. Department of Energy, Office of Science, under Contract No.~DE-AC02-05CH11231.

\bibliography{Hubbard}
\clearpage

\end{document}